\documentstyle[
epsfig]{paper}
\def\lsim{\;\raise0.3ex\hbox{$<$\kern-0.75em\raise-1.1ex\hbox{$\sim$}}\;}
\def\gsim{\;\raise0.3ex\hbox{$>$\kern-0.75em\raise-1.1ex\hbox{$\sim$}}\;}
\def\thebibliography#1{\section*{References}\list
 {[\arabic{enumi}]}{\settowidth\labelwidth{[#1]}\leftmargin\labelwidth
 \advance\leftmargin\labelsep
 \usecounter{enumi}}
 \def\newblock{\hskip .11em plus .33em minus -.07em}
 \sloppy
 \sfcode`\.=1000\relax}
\let\endthebibliography=\endlist

\begin{document}

\title{OVERVIEW OF KAON DECAY PHYSICS}

\author{R. D. Peccei\thanks{Invited talk given at the 23rd INS
International Symposium on Nuclear and Particle Physics with
Meson Beams with 1 GeV/c Region, Tokyo, Japan, March 1995.
To be published in the Symposium Proceedings.}\\
\\
\it Department of Physics, University of California at Los Angeles,\\
 Los Angeles, CA 90095-1547}
\maketitle
\section*{Abstract}

After a brief history of the insights gained from Kaon physics,
the potential of Kaon decays for probing lepton number violation
is discussed.  Present tests of CTP and of Quantum Mechanics in the
neutral Kaon sector are then reviewed and the potential of the
Frascati $\Phi$ factory for doing incisive tests in this area is
emphasized.  The rest of this overview focuses on CP violating effects
in the Kaon system.  Although present observations of CP violation are
perfectly consistent with the CKM model, we emphasize the theoretical
and experimental difficulties which must be faced to establish this
conclusively.  In so doing, theoretical predictions and experimental
prospects for detecting $\Delta S=1$ CP violation through
measurements of $\epsilon^\prime/\epsilon$ and of rare K decays are
reviewed.  The importance of looking for evidence for non-CKM
CP-violating phases, through a search for a non-vanishing transverse
muon polarization in $K_{\mu 3}$ decays, is also stressed.

\section{Introduction}

Ever since their discovery \cite{Brown} nearly 50 years ago,
Kaons have played an important part in the development of particle
physics.  The suggestion of Pais \cite{Pais} and Gell-Mann
\cite{GellMann} that Kaons possessed a new quantum number---
strangeness---and so could only be produced in association
with particles with the opposite quantum number was soon
confirmed experimentally \cite{Fowler} and marked the beginning of
the study of flavor physics.  At the same time, the $\tau-\theta$
puzzle \cite{Dalitz} provided the impetus for Lee and Yang \cite{LY}
to suggest that the weak interactions did not conserve parity.
With parity violation the identification of the $\tau$, which decayed
into three pions, with the $\theta$, which decayed into two pions,
was natural and Kaons were born.

In the 1960's Kaons played an important role in elucidating some of
the underlying symmetries of the strong interactions, well before
the advent of QCD where these symmetries are more manifest.
The approximate invariance of the strong interactions under flavor
SU(3) \cite{GMN} led to the Gell-Mann-Okubo formula \cite{GMO}
interrelating the Kaon mass with that of the pion and the $\eta$.
The extension of this symmetry to a, spontaneously broken,
approximate $SU(3)_{\rm V} \times SU(3)_{\rm A}$ invariance of the
strong interactions \cite{CA} underscored the special dynamical role
of the pseudoscalar meson octet $(\pi,K,\eta)$ as near Nambu
Goldstone bosons \cite{NG}.  It also provided important connections
between weak decay amplitudes involving Kaons, like the Callan-Trieman
relation \cite{CT}.  Almost simultaneously to these theoretical
developments, the discovery of the decay $K_{\rm L}\rightarrow 2\pi$
by Christianson, Cronin, Fitch and Turlay \cite{CCFT} provided the
first indication that CP, like parity, was also not a good symmetry
of nature.

Kaon physics also provided important insights into the flavor structure
of the weak interactions.  The weaker strength of Kaon weak decays
relative to that of the pions lead to the introduction of the Cabibbo
angle \cite{Cabibbo} and to the notion of flavor mixing for charged
current weak interactions.  The very suppressed nature of the
neutral current decay $K_{\rm L}\rightarrow \mu^+\mu^-$, relative
to the charged current decay $K^+\rightarrow \mu^+\nu_e$, found
its natural explanation through the GIM mechanism \cite{GIM}
and lead to the prediction of a further flavor---charm---which was
subsequently found \cite{SPEAR}.

Although perhaps the halcyon decays of Kaon physics are past, Kaons
can be counted on, even today, to provide important future physics
insights at the research frontier.  In this talk I would like to
focus on three such areas, where experiments with Kaon beams can
substantially further our understanding:
\begin{description}
\item [i)] Tests of flavor violation, using the intense Kaon beams now
available, to probe for lepton number violation to an accuracy of
one part in a trillion.
\item [ii)] Tests of CPT and of Quantum Mechanics to unprecedented
accuracy, using to advantage the tiny mass difference between the
$K_{\rm L}$ and the $K_{\rm S}$ states to amplify these
effects and make them experimentally more accessible.
\item [iii)] Tests of CP violation in the only system where this
phenomena has been observed, particularly to look for evidence
for direct $(\Delta S=1)$ CP violation and for CP violation induced by
new scalar interactions.
\end{description}

\section{Testing for Flavor Violation}

Both lepton number (L) and baryon number (B) are classical
global symmetries of the Standard Model.  However, there are no
good reasons why these symmetries should be exact in nature.
In fact, it is known that quantum effects arising from the
existence of chiral anomalies \cite{anomaly} lead to a breakdown
of (B+L)-symmetry \cite{'tHooft}.  Also, if the Standard Model is
embedded into some Grand Unified Theory (GUT), then generally
these theories have lepton and quarks in the same representation,
leading to a breakdown of both L and B \cite{PS}.

The violations of B and L alluded to above are highly
suppressed, leading to phenomena like proton decay which have
extremely long lifetimes \cite{proton}.  However, these may
well not be the only sources of flavor violation in nature.  For
instance, new physics may involve interactions which are mediated by
leptoquarks---objects having both quark and lepton quantum numbers.
Leptoquark exchanges, as those typified by the diagram in Fig. 1,
will give rise to flavor changing decays, like $K_{\rm L}\rightarrow
\mu^+e^-$.

\newpage
\begin{figure}[t!]
{}~\epsfig{file=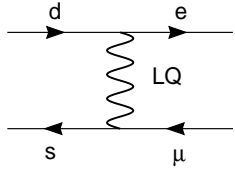,width=13.5cm,height=4cm}
\caption{Leptoquark exchange leading to flavor violation}
\end{figure}

Assuming comparable couplings of the leptoquarks to quarks
and leptons to those of the W's to these excitations, one predicts a
branching ratio for this process of order
\[
BR(K_{\rm L}\rightarrow \mu^+e^-) \sim \left(\frac{M_{\rm W}}
{M_{\rm LQ}}\right)^4
\]
One sees that if one probes flavor violating processes to the
level of ${\rm BR} \sim 10^{-12}$ one is probing leptoquark
masses (and therefore new physics) to the level $M_{\rm LQ} \geq
100~{\rm TeV}$.

Experimentally, searches for lepton flavor violating interactions have
been carried out to great accuracy.  The present best bounds for the
process $K_{\rm L} \rightarrow \mu^+e^-$ come from experiments at
Brookhaven (E791) and KEK (E137).  These experiments have established
90\% CL bounds of $O(10^{-10})$ for this branching ratio:
\begin{eqnarray*}
BR(K_{\rm L}\rightarrow \mu^+e^-) & \leq & 3.3\times 10^{-11}~~~
\mbox{\cite{Arisaka}}\\
BR(K_{\rm L}\rightarrow \mu^+e^-) & \leq & 9.4\times 10^{-11}~~~
\mbox{\cite{Akagi}}
\end{eqnarray*}
A new experiment (E871) has started running at Brookhaven which should
be able to push down the limit for this decay to a $BR \sim 2\times
10^{-12}$.

For the lepton violating process $K \rightarrow \pi\mu e$ the best
bound to date comes from a Brookhaven experiment, E777:
\[
BR(K^+\rightarrow \pi^+\mu^- e^+) \leq 2.1 \times 10^{-10}~~~
\mbox{\cite{Lee}}
\]
A new experiment at Brookhaven (E865) has started running and hopes
to push this BR also down to the level of $O(10^{-12})$.

The large
improvement in precision expected from BNL E865
compared to the present
bound, as well as the sharpening expected from BNL E871 to the
present limit on $K_{\rm L} \rightarrow \mu^+ e^-$, if no effects are
found will produce only a small extension of the mass limits
for particles which could mediate these decays.   Typically,
an improvement
in BR limits of a factor of 10 will only lead to an improvement in
the mass limit by a factor of 2 or so, since the BR scales as
$M^{-4}$.

\section{CPT and Quantum Mechanics Tests}

The CPT theorem \cite{CPT} is a fundamental consequence of being able
to describe elementary particle interactions by a relativistic local
quantum field theory.  Thus violation of CPT invariance would signal the
breakdown of some sacred principles, like locality or even Quantum
Mechanics! Nevertheless, it has been suggested that some small
violation of CPT invariance may possibly arise in connection to
string theory \cite{string} or may result from gravitational effects
\cite{Hawking}.  In both cases a concomitant breakdown of
Quantum Mechanics may also occur.  The neutral Kaon system is ideal
for probing these speculations since the very small $K_{\rm L}-K_{\rm S}$
mass difference allows one to probe $K^o-\bar K^o$ mass differences of
$O(10^{-18})$ the Kaon mass.  This is the right range to begin
seeing possible inverse Planck mass effects:
\[
\frac{M_{K^o}-M_{\bar K^o}}{M_{K^o}} \sim
  \frac{M_{K^o}}{M_{\rm Planck}} \sim 10^{-19}
\]
Present day data are consistent with CPT conservation.  However, more
incisive tests would be welcome.  These are likely to be
carried out in the near future, particularly at the Frascati $\Phi$
Factory.

There have been two kinds of theoretical analyses of CPT
violating phenomena in the neutral Kaon complex which differ in
that in one case \cite{DP} Quantum Mechanics is assumed to hold, while
in the other both CPT and Quantum Mechanics violations are included
\cite{ENM}, \cite{HP}.
If CPT is violated, the phenomenology of the $K^o-\bar K^o$ system
is modified in two ways \cite{DP}:
\begin{description}
\item [i)] The $K_{\rm L}$ and $K_{\rm S}$ states are now different
superpositions of $K^o$ and $\bar K^o$, characterized by separate
mixing parameters $\epsilon_{\rm L}$ and $\epsilon_{\rm S}$:
\begin{eqnarray*}
|K_{\rm L}> & \simeq & \frac{1}{\sqrt{2}}
\{(1+\epsilon_{\rm L})|K^o>-(1-\epsilon_{\rm L})|\bar K^o>\} \\
|K_{\rm S}> & \simeq & \frac{1}{\sqrt{2}}
\{(1+\epsilon_{\rm S})|K^o>+(1-\epsilon_{\rm S})|\bar K^o>\} ~,
\end{eqnarray*}
with
\[
\epsilon_{\rm L} = \epsilon_{\rm K}-\delta_{\rm K}~; \;
\epsilon_{\rm S}=\epsilon_{\rm K}+\delta_{\rm K}
\]
where the parameter $\delta_{\rm K}$ typifies mixing CPT violation.
\item [ii)] Particle and antiparticle decay amplitudes are no longer
simply related by complex conjugation.  Instead one has, for
example\cite{Barmin}:
\begin{eqnarray*}
A(K^o\rightarrow \pi^-\ell^+\nu_e)=a+b & ; &
A(\bar K^o\rightarrow \pi^+\ell^-\bar\nu_e)=a^*-b^* \\
A(K^o\rightarrow 2\pi;I)=(A_I+B_I)e^{i\delta_I} & ; &
A(\bar K^o\rightarrow 2\pi;I)=(A_i^*-B_I^*)e^{i\delta_I}
\end{eqnarray*}
In the above, the $b$ and $B_I$ amplitudes violate CPT.
\end{description}

If, in addition, also Quantum Mechanics is violated then,
besides the above modifications due to CPT non-invariance,
the time evolution of the $K^o-\bar K^o$ complex is different
from the usual Schr\"odinger evolution.  This is most easily
described in terms of the evolution of the density matrix $\rho$
of the $K^o-\bar K^o$ system.  Quantum Mechanics violation is
introduced \cite{Hawking} through the appearance of an extra
term in the Schr\"odinger equation for $\rho$ \footnote{For the
$K^o-\bar K^o$ system the $2\times 2$ Hamiltonian $H$ is not
Hermitian, since it also describes the decay of these states:
$H=M-\frac{i}{2}\Gamma$.}
\[
i\frac{\partial}{\partial t}\rho =
H\rho-\rho H^\dagger +\delta h\rho~.
\]
Because of the presence of the $\delta h$ term above, the evolution
of $\rho$ with time has no longer the Schr\"odinger form.
Given $\delta h$ this evolution can be straightforwardly
computed \cite{ENM},\cite{HP}.  Ellis {\it et al.} \cite{EHNS}
show that the simplest $\delta h$, which is consistent with
some general principles like probability conservation, can be
parametrized by three CPT and Quantum Mechanics violating
parametes: $\alpha,\beta,\gamma$, with
\[
\alpha,\gamma > 0~; \; \alpha\gamma > \beta^2~.
\]

Present day data is not sufficient to determine all these CPT violating
parameters.  In addition, for the case where one assumes that
there is also a violation of Quantum Mechanics, one should
really do a fit of the experimental data with the modified
evolution equation.  Without a violation of Quantum Mechanics,
there are basically two independent tests of CPT: one where the $K_{\rm L}$
semileptonic decay asymmetry $A_{\rm K_L}$ is compared to the real
part of $\eta_{+-}$, and the other where the phase of $\epsilon$ is
compared to the superweak phase $\phi_{\rm SW}$ \cite{DP}.\footnote
{For these tests, given the present accuracy and the smallness of
$\epsilon^\prime$, one can neglect $\epsilon^\prime$ altogether.  Thus
one has $\eta_{+-}\simeq \eta_{oo} \simeq \epsilon$.}  The first test
is sensitive to amplitude CPT violation and one has
\[
\frac{1}{2}A_{\rm K_L}-Re~\eta_{+-}=
\frac{Re~b}{Re~a}-\frac{Re~B_o}{Re~A_o}=
(1.3 \pm 6.6)\times 10^{-5}~,
\]
where the right-hand side uses the PDG values \cite{PDG} for
the experimental quantities.  For the second test one uses the
fact that one can decompose $\epsilon$ into a CP violating and a CPT
violating piece, with these terms being $90^\circ$ out of
phase \cite{Cronin}:
\[
\epsilon=\epsilon_{CP \!\!\!\!/} e^{i\phi_{\rm SW}} +
\epsilon_{CP \!\!\!\!/ T} e^{i\left(\phi_{\rm SW}+\frac{\pi}{2}\right)}
\]
with
\[
\phi_{\rm SW} = \tan^{-1}
\frac{2\Delta m}{\Gamma_{\rm S}-\Gamma_{\rm L}}=
(43.64\pm 0.14)^\circ~.
\]
One finds, again using PDG values, that
\[
\epsilon_{CP \!\!\!\!/ T} \simeq \sqrt{2}~Im~\delta_{\rm K} \simeq \sqrt{2}
\left(\frac{Re~B_o}{Re~A_o}-Re~\delta_{\rm K}\right)=
(2.6\pm 3.2)\times 10^{-5}~,
\]
so that $\epsilon_{CP \!\!\!\!/ T}$ is at most of order 1\% of
$\epsilon$.  Hence the decay $K_{\rm L}\rightarrow 2\pi$ is either
wholly, or predominantly, a result of CP violation, not CPT
violation.  Nevertheless, because what one measures are
essentially differences of CPT violating parameters, one cannot
exclude an accidental cancellation and thus the possibility of
having a hidden large CPT violation \cite{DP}.
If amplitude CPT violation is
neglected, then this cancellation is excluded and a measurement
of $\epsilon_{CP \!\!\!\!/ T}$ at the level indicated above gives
a strong bound on the $K^o-\bar K^o$ mass difference:
\[
M_{K^o}-M_{\bar K^o} \simeq 2\sqrt{2}~\Delta m
\epsilon_{CP \!\!\!\!/ T}=(2.57 \pm 3.18)\times 10^{-19}~{\rm GeV}~.
\]

Huet and Peskin \cite{HP} have recently performed an analysis of
the time evolution of the decay of an initial $K^o$ into $\pi^+\pi^-$,
under the assumption that Quantum Mechanics is violated by the
$\delta h$ perturbation discussed above.  Such decays are studied
in the CP Lear experiment \cite{Adler}, since in $p\bar p$
annihilation one can tag the produced $K^o$ or $\bar K^o$ states
with the sign of the accompanying produced charged Kaon.
Neglecting amplitude CPT
violation, the decay of an initial $K^o$ into $\pi^+\pi^-$ can be
written as
\[
R_{+-}(t)=e^{-\Gamma_{\rm S}t}+R_{\rm L} e^{-\Gamma_{\rm L}t}
+2|\epsilon^-_{\rm L}|\cos (\Delta mt+\phi_{+-})
\exp\left[-\frac{(\Gamma_{\rm L}+\Gamma_{\rm S})t}{2}\right]~.
\]
If there is no violation of Quantum Mechanics, then
\[
|\epsilon^-_{\rm L}|=|\eta_{+-}|~; \;\; R_{\rm L}=|\eta_{+-}|^2~.
\]
If Quantum Mechanics is violated, however, $R_{\rm L}$ and
$\epsilon^-_{\rm L}$ are no longer simply related and they
depend on the CPT violating parameters $\beta$ and $\gamma$\footnote
{The parameter $\alpha$ affects the precise exponential decrease in
the above equation.  However, this change can be neglected
in the analysis\cite{HP}.}.
One finds:
\[
\epsilon^-_{\rm L}=\epsilon_{\rm L}-\frac{\beta}{d}~; \;\;
R_{\rm L}=|\epsilon^-_{\rm L}|^2+\frac{\gamma}{\Delta\Gamma}+
\frac{4\beta}{\Delta\Gamma}~{\rm Im}~ \frac{\epsilon^-_{\rm L}d}
{d^*}~,
\]
where the kinematical parameter $d$ is
\[
d=\Delta m+\frac{i}{2}(\Gamma_{\rm S}-\Gamma_{\rm L})\equiv \Delta m+
\frac{i}{2}\Delta\Gamma \simeq (5\times 10^{-15}{\rm GeV})
e^{i\phi_{\rm SW}}~.
\]

By comparing the time evolution $R_{+-}(t)$ observed by CP Lear
\cite{Adler} with their expression, Huet and Peskin \cite{HP} are
able to
extract values for the parameters $\beta$ and $\gamma$.  Interestingly,
even assuming that there are Quantum Mechanics violations, one can
attribute at most only 10\% of $\epsilon$ to CPT violation.  So, even in
this more extreme scenario, the measurement of a nonvanishing value for
$\epsilon$ is principally, or exclusively, a signal of CP violation.
I quote below the results obtained by Huet and Peskin {\cite{HP} when
also amplitude CPT violation is included.
They find \footnote{These results if
$\beta=\gamma=0$ give looser bounds on the CPT violating amplitude
combination $\frac{{\rm Re}~b}{{\rm Re}~a}-\frac{\rm {Re~B_o}}
{\rm {Re~A_o}}$ than what was quoted above, since only the CP Lear data
was used.  From the rate determination one has a value of
$(3.3\pm 10.3)\times 10^{-5}$ for the CPT violating amplitude
combination, while this becomes $(4.8 \pm 15.9)\times 10^{-5}$
from the interference determination.}:
\[
\beta + \frac{|d|}{2\sin\phi_{\rm SW}}
\left[\frac{{\rm Re}~b}{{\rm Re}~a}-\frac{{\rm Re}~B_o}{{\rm Re}~A_o}\right]
=(1.2 \pm 4.4)\times 10^{-19}~{\rm GeV}
\]
\[
\gamma-2|\eta_{+-}d|
\left[\frac{{\rm Re}~b}{{\rm Re}~a}-
\frac{{\rm Re}~B_o}{{\rm Re}~A_o}\right]=
(-1.1\pm 3.6)\times 10^{-21}~{\rm GeV}~.
\]
The parameter $\beta$ also gives a contribution to $\epsilon$ at
$90^\circ$ to $\phi_{\rm SW}$.  So, if Quantum Mechanics is violated,
the phase difference of the phase of $\epsilon \simeq \eta_{+-}$
from $\phi_{\rm SW}$ now not only measures $\epsilon_{\rm CP\!\!\!\!/T}$
but $\epsilon_{\rm CP\!\!\!\!/T}-\frac{\sqrt{2}\beta}{|d|}$.  Using
the PDG values for the difference between $\phi_{+-}$ and
$\phi_{\rm SW}$ one finds the additional constraint:
\[
\beta-\frac{|d|}{\sqrt{2}}\epsilon_{\rm CP\!\!\!\!/T}=
(-0.9\pm 1.1) \times 10^{-19}~{\rm GeV}~.
\]

One must do more than just study $K_{\rm L}$ semileptonic decays and
$K_{\rm S}$ and $K_{\rm L}$ decays into 2 pions to distinguish all
of the parameters connected with possible CPT and Quantum Mechanics
violations.  The $\Phi$ factory presently under construction at
Frascati is ideal for this task, although already some important
new information should emerge from CP Lear.  Indeed, we learned at this
meeting \cite{Lear} that CP Lear has a preliminary measurement of the
$K_{\rm S}$ semileptonic symmetry $A_{\rm K_S}$ which agrees with
$A_{\rm K_L}$ within 10\%.  If Quantum Mechanics is OK, one expects
\[
A_{\rm K_S}-A_{\rm K_L}=-4{\rm Re}~\delta_{\rm K}
\]
and such a measurement isolates ${\rm Re}~\delta_{\rm K}$
directly.

At a $\Phi$ factory one can perform CPT and Quantum Mechanics
tests principally by using the accelerator as a $K^o-\bar K^o$
interferometer.  Additionally, one can use $K_{\rm L}$ decays as
a tag to study $K_{\rm S}$ decays and perform tests of the type
described above.  The initial state produced at a $\Phi$ factory,
when the $\Phi$ decays, is a coherent superposition of $K_{\rm S}$
and $K_{\rm L}$ states:
\[
|\Phi>=\frac{1}{\sqrt{2}}
\{|K_{\rm S}(\vec p)\rangle|K_{\rm L}(-\vec p)\rangle
-|K_{\rm S}(-\vec p)\rangle|K_{\rm L}(\vec p)\rangle\}~.
\]
As a result, when the $K_{\rm S}$ and $K_{\rm L}$ states
eventually decay into final states $f_1$ and $f_2$, the relative
time decay probability will show a characteristic interference
pattern reflecting the initial coherent superposition.
This interference pattern is sensitive to CP and CPT violation
parameters \cite{Lipkin}.  For instance, if the final states $f_1$
and $f_2$ are $\pi^+\pi^-$ and $\pi^o\pi^o$, for large time
differences between the times $t_1$ and $t_2$ where the $\pi^+\pi^-$
and $\pi^o\pi^o$ are produced, the decay probability will fall as
$e^{-\Gamma_L|t_1-t_2|}$.  However, the coefficient of this
exponential is different depending on whether $t_1 \gg t_2$ or
$t_2 \gg t_1$, with this difference being related to ${\rm Re}~
\epsilon^\prime/\epsilon$ \cite{DAFNE}.

In the case Quantum Mechanics is violated, these interference
patterns will be altered.  By studying in detail the time
evolution of the system one should then be able to separate out pure
effects of CPT violation from effects in which both CPT and Quantum
Mechanics are violated.  A nice example to study \cite{HP} is the
pattern of the time evolution for identical final states ($f_1=f_2$).
Because of the antisymmetry in the initial $K_{\rm L},K_{\rm S}$
state, it is easy to see that quantum mechanically the decay
probability vanishes if the decays into $f_1$ and $f_2$ occur at
precisely the same time ($t_1=t_2$).  This is no longer the case when
one admits possible Quantum Mechanics violations.  For example, if
$f_1$ and $f_2$ are both semileptonic states, one has \cite{HP}:
\begin{eqnarray*}
&&I\bigl(\ell^\pm\pi^\mp\nu_{\ell}(t_1);~\ell^\pm \pi^\mp\nu_\ell (t_2)\bigr)
 =\Bigl\{(1\pm 4~{\rm Re}~\epsilon_{\rm K})
  \biggl[e^{-\Gamma_{\rm S}t_1-\Gamma_{\rm L}t_2}  \\
 & + &e^{-{\Gamma_{\rm S}t_2-\Gamma_{\rm L}t_1}}-2\cos \Delta m(t_1-t_2)
  \exp -\bigl(\frac{\Gamma_{\rm S}+\Gamma_{\rm L}}{2}
  +\alpha-\gamma\bigr)(t_1+t_2)\biggr] \\
 & \pm & \frac{4\beta}{|d|}\biggl[\Bigl[\sin(\Delta mt_1-\phi_{\rm SW})
  \exp -\bigl[\frac{\Gamma_{\rm S}+\Gamma_{\rm L}}{2}
  +\alpha-\gamma\bigr]t_1~ e^{-\Gamma_{\rm S}t_2}+
  (t_1\leftrightarrow t_2)\Bigr] \\
  & + & [\Gamma_{\rm S}\leftrightarrow \Gamma_{\rm L};~
  \phi_{\rm SW} \leftrightarrow -\phi_{\rm SW}]\biggr] \\
 & + &\frac{2\alpha}{\Delta m}\sin \Delta m(t_1+t_2)
  \exp -\bigl(\frac{\Gamma_{\rm S}+\Gamma_{\rm L}}{2}
  +\alpha-\gamma\bigr)(t_1+t_2) \\
 & + &\frac{2\gamma}{\Gamma_{\rm S}-\Gamma_{\rm L}}
 \left[e^{-\Gamma_{\rm L}(t_1+t_2)}-e^{-\Gamma_{\rm S}(t_1+t_2)}
  \right]\Bigr\}~.
\end{eqnarray*}
The first term in the curly bracket, if $\alpha-\gamma=0$, is the
usual quantum mechanical expression which vanishes when $t_1=t_2$.
The others three terms are proportional to the additional parameters
$\alpha,\beta$ and $\gamma$ connected with Quantum Mechanics violation.
Because the time dependence of all these four terms is
different, in principle
a careful study of this quantity would allow a separate determination
of $\alpha$, $\beta$, $\gamma$ and ${\rm Re}~\epsilon_{\rm K}$.

\section{CP Violation}

To date the neutral Kaon system is the only place where a violation
of CP has been observed.\footnote{This is not quite correct, since
to obtain a non-trivial asymmetry between matter and antimatter
in the universe, it is necessary that there should be processes that
violate CP\cite{Sakharov}.}  In the modern gauge theory paradigm
this phenomena can have one of two possible origins.  Either
\begin{description}
\item[i)] the full Lagrangian of the theory is CP invariant, but this
symmetry is not preserved by the vacuum state: CP $|0\rangle \not=
|0\rangle$.  In this case CP is a spontaneously broken
symmetry \cite{TDLee}.
\end{description}
or
\begin{description}
\item[ii)] there are terms in the Lagrangian of the theory which
are not invariant under CP transformations.  CP is explicitly broken
by these terms and is no longer a symmetry of the theory.
\end{description}

The first possibility, unfortunately, runs into a potential
cosmological problem\cite{KOZ}.  As the universe cools below a
temperature $T^*$ where spontaneous CP violation occurs, one
expects that domains of different CP should form.  These domains
are separated by walls having a typical surface energy density
$\sigma\sim T^{*^3}$.  The energy density associated with these walls
dissipates slowly as the universe cools further and eventually
contributes an energy density to the universe at temperature T of order
$\rho_{\rm Wall}\sim T^{*^3}T$.  Such an energy density today would
typically exceed the universe closure density by many orders of
magnitude:
\[
\rho_{\rm Wall}\sim 10^{-7}\left(\frac{T^*}{\rm TeV}\right)^3
{\rm GeV}^{-4} \gg \rho_{\rm closure} \sim 10^{-46}~{\rm GeV}^{-4}~.
\]
One can avoid this difficulty by imaging that the scale where CP is
spontaneously violated is very high, so that $T^*$ is above the
temperature where inflation occurs.  In this case the problem
disappears, since the domains get inflated anyway.  Nevertheless,
there are still problems, since it proves difficult to connect
this high energy spontaneous breaking of CP with the observed
phenomenon at low energies.  What emerges, in general, are models
which are quite complex\cite{Barr}, with CP violation being associated
with new interactions much as in the original superweak model of
Wolfenstein\cite{superweak}.

If, on the other hand, CP is explicitly broken the phenomenology of
neutral Kaon CP violation is a quite
natural result of the standard
model of the electroweak interactions.  There is, however, a requirement
emerging from the demand of renormalizability which bears mentioning.
Namely, if CP is explicitly broken then renormalizability requires that
all the parameters in the Lagrangian which can be complex must be
so.  A corollary of this is that the number of possible CP violating
phases in the theory increases with the complexity of the theory, as there
are then more terms which can have imaginary coefficients.

In this respect, the three generation ($N_g=3$) standard model with
only one Higgs doublet is the simplest possible model, since it
has only one phase.  With just one Higgs doublet,
the Hermiticity of the scalar potential allows
no complex parameters to appear.  If CP
is not a symmetry, complex Yukawa couplings are, however, allowed.
After the breakdown of the $SU(2)\times U(1)$ symmetry, these couplings
produce complex mass matrices.  Going to a physical basis with real
diagonal masses introduces a complex mixing matrix in the charged
currents of the theory.  For the quark sector, this is the famous
Cabibbo-Kobayashi Maskawa (CKM) matrix\cite{CKM}.\footnote{If the
neutrinos are massless, there is no corresponding matrix in the
lepton sector since it can be reabsorbed by redefining the neutrino
fields.}  This $N_g\times N_g$ unitary matrix contains $N_g(N_g-1)/2$
real angles and $N_g(N_g+1)/2$ phases.  However, $2N_g-1$ of these phases
can be rotated away by redefinitions of the quark fields leaving only
$(N_g-1)(N_g-2)/2$ phases.  Thus for $N_g=3$ the standard model has
only one physical complex phase to describe all CP violating
phenomena.\footnote{This is not quite true.  In the SM there is also another
phase related to the QCD vacuum angle which leads to a CP violating
interaction involving the gluonic field strength and its dual:
\[
{\cal{L}}_{\rm CP~viol.}=\bar\theta \frac{\alpha_{\rm S}}{8\pi}
F_a^{\mu\nu}\tilde F_{a\mu\nu}~.
\]
The phase angle $\bar\theta$ contributes to the neutron electric
dipole moment and, to respect the existing bound on
$d_n$ \cite{PDG} must be extremely small:
$\bar\theta \leq 10^{-9}-10^{10}$.
Why this should be so is unknown and constitutes
the strong CP problem \cite{strongCP}.}

If CP is broken explicitly, it follows by the renormalizability
corollary that any extensions of the SM will involve further CP
violating phases.  For instance, if one has two Higgs doublets,
$\Phi_1$ and $\Phi_2$, then the Hermiticity of the scalar potential
no longer forbids the appearance of complex terms like
\[
V = \ldots \mu_{12}\Phi_1^\dagger \Phi_2+\mu_{12}^*\Phi_2^\dagger\Phi_1~.
\]
Indeed, if one did not include such terms, the presence of complex
Yukawa couplings would induce such terms at one loop.\footnote{More
precisely, one needs complex counterterms to absorb the complex
quadratic divergences induced through the Yukawa couplings.}

\subsubsection*{Testing the CKM paradigm}

One does not really know if the complex phase present in the CKM
matrix is responsible for the CP violating phenomena observed
in the neutral Kaon system.  Indeed, one does not know whether
there are further phases besides the CKM phase.  Nevertheless, it is
remarkable that, as a result of the hierarchial structure of the
CKM matrix and of other dynamical circumstances, one can {\bf qualitatively}
explain all we know experimentally about CP violation today on the
basis of the CKM picture.
In what follows, I make use of the CKM matrix in the
parametrization adopted by the PDG\cite{PDG} and expand the three
real angles in the manner suggested by Wolfenstein\cite{Wolf} in
powers of the sine of the Cabibbo angle $\lambda$.  To order $\lambda^3$
one has then
\[
V=\left|
\begin{array}{ccc}
1-\frac{\lambda^2}{2} & \lambda & A\lambda^3(\rho-i\eta) \\
-\lambda & 1-\frac{\lambda^2}{2} & A\lambda^2 \\
A\lambda^3(1-\rho-i\eta) & -A\lambda^2 & 1
\end{array}
\right|
\]
with $A,\rho$ and $\eta$ being parameters one needs to fix from
experiments---with $\eta \not= 0$ signalling CP violation.\footnote{It
is often convenient instead of using $\rho-i\eta$ to write this in terms
of a magnitude and phase: $\rho-i\eta=\sigma e^{-i\delta}$, with $\delta$
being the CP violating CKM phase.}

Three pieces of experimental data provide today independent dynamical
information on CP violation.  These are:
\begin{description}
\item[i)] The value of the $K_{\rm L}$ to $K_{\rm S}$ amplitude
ratios, $n_{+-}=\epsilon+\epsilon^\prime$; $\eta_{oo}=\epsilon-
2\epsilon^\prime$, with
\[
|\eta_{+-}|\simeq |\eta_{oo}| \sim 2\times 10^{-3}~.
\]
\item[ii)] The small value of the $\Delta S=1$ CP violating parameter
$\epsilon^\prime$, with the ratio
\[
\epsilon^\prime/\epsilon \lsim 10^{-3}~.
\]
\item[iii)] The very strong bounds on the electric dipole moments
of the neutron and the electron
\[
d_e,d_n \lsim 10^{-25}~{\rm ecm}~.
\]
\end{description}
Other information at hand is either too insensitive, like the
corresponding CP violating parameters for $K\rightarrow 3\pi$ decays
$\eta_{+-o}$ and $\eta_{ooo}$, or is dynamically fixed, like
$A_{\rm K_L}=2~{\rm Re}~\eta_{+-}$ or $\phi_{+-}=\phi_{\rm SW}$
which follows as
a result of CPT invariance.

One can ``understand" the above three facts quite simply in the
CKM paradigm.  In the model $\eta_{+-}$, or the parameter $\epsilon$,
is determined by the ratio of the imaginary to the real part of the
box graph of Fig. 2a.  It is easy to check that this ratio is of
order
\[
\epsilon \sim \lambda^4 \sin\delta \sim 10^{-3} \sin\delta~.
\]
That is, $\epsilon$ is naturally small because of the suppression of
interfamily mixing without requiring the CKM phase $\delta$ to be
small.  Similarly, one can qualitatively understand why $\epsilon^\prime/
\epsilon$ is small.  This ratio is suppressed by the $\Delta I=1/2$ rule
and it is induced only by the Penguin diagrams of Fig. 2b involving
the emission of virtual gluons (or photons), which are Zweig
suppressed\cite{GW}.  Typically this gives
\[
\frac{\epsilon^\prime}{\epsilon} \sim \frac{1}{20} \cdot
\left[\frac{\alpha_{\rm S}}{12\pi}\ln \frac{m_t^2}{m_c^2}\right]
\sim 10^{-3}~.
\]
Finally, in the CKM model the electric dipole moments are small since
the first nonvanishing contributions\cite{Shabalin} occur at three
loops, as shown in Fig. 2c, leading to the estimate\cite{edm}
\[
d_{\rm q,e} \sim em_{\rm q,e}
\left[\frac{\alpha^2 \alpha_{\rm S}}{\pi^3}\right]
\left[\frac{m_t^2 m_b^2}{M_{\rm W}^6}\right] \lambda^6 \sin
\delta \sim 10^{-32}~{\rm ecm}~.
\]

\newpage
\begin{figure}[t!]
{}~\epsfig{file=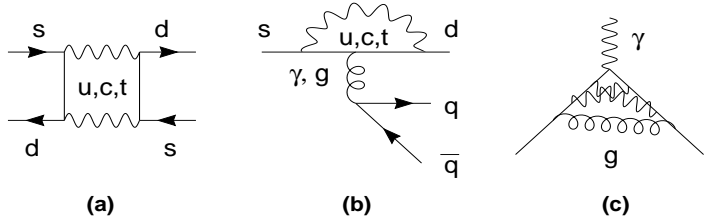,width=13.5cm,height=5cm}
\caption{Graphs contributing to $\epsilon,\, \epsilon'$ and $d_{\rm q,e}$}
\end{figure}

One can, of course, use the precise value of $\epsilon$ measured
experimentally to determine an allowed region for the parameters
entering in the CKM matrix.  Because of theoretical uncertainties
in evaluating the hadronic matrix element of the $\Delta S=2$
operator associated with the box graph of Fig. 2a this parameter
space region is rather large.  Further restrictions on the allowed
values of CKM parameters come from semileptonic B decays and from
$B_d-B_{\bar d}$ mixing.  Because the parameter $A$, related to
$V_{cb}$, is better known, it has become traditional to present the
result of these analysis as a plot in the $\rho-\eta$ plane.  Fig. 3
shows the results of a recent analysis, done in collaboration with my
student, K. Wang\cite{PW}.  The input parameters used, as well as the
range allowed for certain hadronic amplitudes and other CKM matrix
elements is detailed in Table. \ref{RhoParam}

\begin{table}[h!]
\caption[]{
\label{RhoParam}
Parameters used in the $\rho-\eta$ analysis of \cite{PW}
}
\begin{eqnarray*}
\begin{array}{rclr}
|\epsilon| & = & (2.26 \pm 0.02)\times 10^{-3} & \mbox{~~~~~~~\cite{PDG}} \\
\Delta m_d & = & (0.496 \pm 0.032) ps^{-1} & \mbox{\cite{Forty}} \\
m_t & = & (174 \pm 10^{+13}_{-12})~{\rm GeV} & \mbox{\cite{CDF}} \\
|V_{cb}| & = & 0.0378 \pm 0.0026 & \mbox{\cite{Stone}} \\
|V_{ub}|/|V_{cb}| & = & 0.08 \pm 0.02 & \mbox{\cite{Stone}} \\
B_{\rm K} & = & 0.825 \pm 0.035 & \mbox{\cite{Sharpe}} \\
\sqrt{B_d}~f_{B_d} & = & (180 \pm 30)~{\rm MeV} & \mbox{\cite{Lattice}}
\end{array}
\end{eqnarray*}
\end{table}

The resulting $1\sigma$ allowed contour emerging from the overlap
of the three constraints coming from $\epsilon$, $B_{\rm d}-B_{\rm \bar d}$
mixing and the ratio of $|V_{ub}|/|V_{cb}|$, shown in Fig. 4, gives a roughly
symmetric
region around $\rho=0$ within the ranges
\[
0.2 \leq \eta \leq 0.5~; \;\; -0.4 \leq \rho \leq 0.4~.
\]
As anticipated by our qualitative discussion this region implies
that the CKM phase $\delta$ is large ($\rho=0$ corresponds to
$\delta=\pi/2$).  One should note, however, that this analysis
does not establish the CKM paradigm.  Using only the B physics
constraints one sees that in Fig. 3 there is also an overlap
region for $\eta=0$, which gives $\rho=-0.33 \pm 0.08$ \cite{PW}.
So one can still imagine that $\epsilon$ is due to some other CP
violating interaction, as in the superweak model \cite{superweak},
with the CKM phase $\eta$ being very small.  Obviously, it is important
to exclude such a possibility, but this is not going to be easy.  Wang
and I \cite{PW} discussed how this could perhaps happen as a result of
improving the bounds on $B_{\rm s}-B_{\rm\bar s}$ mixing.  Here I would
like to concentrate on what can the Kaon system tell us on this issue.

\begin{figure}[t!]
{}~\epsfig{file=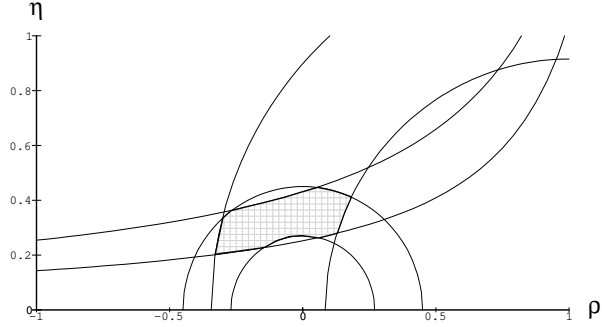,width=13.5cm,height=5cm}
\caption{Constraints on the $(\rho, \, \eta)$ plot}
\end{figure}

\begin{figure}[t]
{}~\epsfig{file=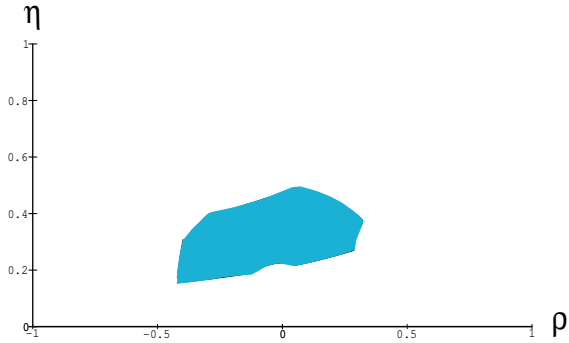,width=13.5cm,height=5cm}
\caption{Allowed region in the $\rho \, - \, \eta$ plane}
\end{figure}

In principle, one can obtain quantitative tests of the CKM model with
Kaon experiments.  However, the needed experiments are very challenging,
either due to the high precision required or due to the rarity of the
processes to be studied.  Furthermore, the analysis of these results
is also theoretically very challenging, since it will require
better estimates of hadronic matrix elements than what we have at
present.

A good example of both of these challenges is provided by
$\epsilon^\prime/\epsilon$.  The present data on this ratio is
inconclusive, with the result obtained at Fermilab \cite{E731}.
\[
{\rm Re}~\frac{\epsilon^\prime}{\epsilon}=
(7.4 \pm 5.2 \pm 2.9) \times 10^{-4} ~~~\mbox{[E731]}
\]
being consistent with zero within the error, while the result
of the NA31 experiment at CERN \cite{NA31} giving a non-zero value
to $3\sigma$:
\[
{\rm Re}~\frac{\epsilon^\prime}{\epsilon}=
(23.0 \pm 3.6 \pm 5.4) \times 10^{-4} ~~~\mbox{[NA31]}
\]
Theoretically, the predictions for $\epsilon^\prime/\epsilon$ are
dependent both on the value of the CKM matrix elements and, more
importantly, on an estimate of certain hadronic matrix elements.
Buras and Lautenbacher \cite{BL} give for this ratio the approximate
formula
\[
{\rm Re}~\frac{\epsilon^\prime}{\epsilon} \simeq
3.6\times 10^{-3} A^2\eta
\left[B_6-0.175\left(\frac{m^2_t}{M_{\rm W}^2}\right)^{0.93}B_8\right]~.
\]
Here $B_6$ and $B_8$ are quantities related to the matrix elements
of the dominant gluonic and electroweak Penguin operators,
respectively.  The electroweak Penguin contribution is suppressed
relative to the gluonic Penguin contribution by a factor of
$\alpha/\alpha_{\rm S}$.  However, it does not suffer from the
$\Delta I=3/2$ suppression and so one gains back a factor of 20.
Furthermore, as Flynn and Randall \cite{FR} first noted, the
contribution of these terms can become significant for large top
mass because it grows approximately as $m_t^2$.  The result of the CKM
analysis presented earlier suggested that
\[
0.12 \leq A^2\eta \leq 0.31~,
\]
while for $m_t=175~{\rm GeV}$ the square bracket above reduces to
[$B_6-0.75B_8$].  Hence one can write the expectation from theory
for $\epsilon^\prime/\epsilon$ as
\[
4.3 \times 10^{-4}[B_6-0.75B_8] \leq
{\rm Re}~\frac{\epsilon^\prime}{\epsilon} \leq 11.2 \times 10^{-4}
[B_6-0.75B_8]~.
\]
Because the top mass is so large, the predicted value for
$\epsilon^\prime/\epsilon$ depends rather crucially on {\bf both} $B_6$
and $B_8$.  These (normalized) matrix elements have been estimated
by both lattice \cite{Ciuchini} and $1/N$ \cite{N} calculations
to be equal to each other, with an individual error of $\pm 20\%$:
\[
B_6=B_8=1 \pm 0.20~.
\]
Thus, unfortunately, the combination entering in $\epsilon^\prime/
\epsilon$ is poorly known.  It appears that the best
one can say theoretically
is that ${\rm Re}~\epsilon^\prime/\epsilon$ should be a ``few" times
$10^{-4}$, with a ``few" being difficult to pin down more precisely.
Theory, at any rate, seems to favor the E731 experimental result
over that of NA31.

Fortunately, we may learn something more in this area in the next
five years or so.  There are 3rd generation experiments in preparation
both at Fermilab (KTeV) and CERN (NA48).  These experiments should
begin taking data in a year or so and are designed to reach statistical
and systematic accuracy for $\epsilon^\prime/\epsilon$ at the level
of $10^{-4}$.  The Frascati $\Phi$ factory which should begin
operations in 1997, in principle, can also provide interesting information
for $\epsilon^\prime/\epsilon$.  At the $\Phi$ factory one will need
an integrated luminosity of $\int {\cal{L}}~dt = 10~fb^{-1}$ to
arrive at a statistical sensitivity for $\epsilon^\prime/\epsilon$
at the level of $10^{-4}$.  However, if this statistical sensitivity
is reached, the systematic uncertainties will be quite different than
those at KTeV and NA48, providing a very useful cross check.  It is
important to remark that, irrespective of detailed theoretical
prediction, the observation of a non-zero value for $\epsilon^\prime/
\epsilon$ at a significant level is very important, for it would
provide direct evidence for $\Delta S=1$ CP violation and would rule
out a superweak explanation for the observed CP violation in the
neutral K sector.

\subsubsection*{Rare Kaon Decays}

There are alternatives to the $\epsilon^\prime/\epsilon$ measurement
which could reveal $\Delta S=1$ (direct) CP violation.  However, these
alternatives involve daunting experiments\cite{RW},
which are probably out of reach
in the near term.  Whether these experiments can (or will?)
eventually be carried out is an open question which I will return to
later.

\subsection*{$K_{\rm S}$ decays}

CP Lear already and the Frascati $\Phi$ factory soon will enable a
more thorough study of $K_{\rm S}$ decays by more efficient
tagging.  The decay $K_{\rm S}-3\pi^o$ is CP-violating, while
the $K_{\rm S}\rightarrow \pi^+\pi^-\pi^o$ mode has both CP-conserving
and CP-violating pieces.  However, even in this case the CP
conserving piece is small and vanishing in the center of the Dalitz
plot.  Hence one can extract information about CP violation from
$K_{\rm S}\rightarrow 3\pi$ decays.  The analogue $K_{\rm S}/K_{\rm L}$
amplitude ratios to $\eta_{+-}$ and $\eta_{oo}$ for $K\rightarrow 3\pi$
have both $\Delta S=1$ and $\Delta S=2$ pieces:
\[
\eta_{ooo}=\epsilon+\epsilon^\prime_{ooo}~; \;\;
\eta_{+-o}=\epsilon+\epsilon^\prime_{+-o}~.
\]
However,
in contrast to what obtains in the $K\rightarrow 2\pi$ case, here
the $\Delta S=1$ pieces can be larger.  Cheng \cite{Cheng} gives
estimates for $\epsilon^\prime_{+-o}/\epsilon$ and
$\epsilon^\prime_{ooo}/\epsilon$ of $O(10^{-2})$, while others
are more pessimistic \cite{pessimistic}.  Even so, there does not
appear to be any realistic prospects in the near future
to probe for $\Delta S=1$
CP-violating amplitudes in $K_{\rm S}\rightarrow 3\pi$.  For instance, at
a $\Phi$-factory even with an integrated luminosity of $10~{\rm fb}^{-1}$
one can only reach an accuracy for $\eta_{+-o}$ and $\eta_{ooo}$
of order $3\times 10^{-3}$, which is at the level of $\epsilon$ not
$\epsilon^\prime$.

\subsection*{Asymmetries in charged K decays}

CP violating effects involving charged Kaons can only be due to
$\Delta S=1$ transitions, since $K^+\leftrightarrow K^-$ $\Delta S=2$
mixing is forbidden by charge conservation.  A typical CP-violating
effect in charged Kaon decays necessitates a comparison between
$K^+$ and $K^-$ processes.  However, a CP-violating asymmetry between
these processes can occur only if there are at least two decay
amplitudes involved and these amplitudes have both a relative weak
CP-violating phase and a relative strong rescattering phase between
each other.  Thus the resulting asymmetry necessarily depends on
strong dynamics.  To appreciate this fact, imagine writing the
decay amplitude for $K^+$ decay to a final state $f^+$ as
\[
A(K^+\rightarrow f^+)=A_1~e^{i\delta_{\rm W_1}}e^{i\delta_{\rm S_1}}+
A_2~e^{i\delta_{\rm W_2}}e^{i\delta_{\rm S_2}}~.
\]
Then the corresponding amplitude for the decay $K^-\rightarrow f^-$
is
\[
A(K^-\rightarrow f^-)=A_1~e^{-i\delta_{\rm W_1}}e^{i\delta_{S_1}}+
A_2~e^{-i\delta_{\rm W_2}}e^{i\delta_{\rm S_2}}~.
\]
That is, the strong rescattering phases are the same but one complex
conjugates the weak amplitudes.  From the above one sees that the
rate asymmetry between these processes is
\begin{eqnarray*}
{\cal{A}}(f^+;f^-)& = & \frac{\Gamma(K^+\rightarrow f^+)-
\Gamma(K^+\rightarrow f^-)}{\Gamma(K^+\rightarrow f^+)+
\Gamma(K^-\rightarrow f^-)} \\
& = & \frac{2A_1A_2\sin(\delta_{\rm W_2}-\delta_{\rm W_1})
\sin(\delta_{\rm S_2}-\delta_{\rm S_1})}
{A_1^2+A_2^2+2A_1A_2\cos(\delta_{\rm W_2}-\delta_{\rm W_1})
\cos(\delta_{\rm S_2}-\delta_{\rm S_1})}~.
\end{eqnarray*}

Unfortunately,
detailed calculations in the standard CKM paradigm for rate
asymmetries and asymmetries in Dalitz plot parameters for various
charged Kaon decays give quite tiny predictions.  This can be
qualitatively understood as follows.  The ratio
$A_2 \sin(\delta_{\rm W_2}
-\delta_{\rm W_1})/A_1$ is typically that of a Penguin amplitude to
a weak decay amplitude and so is of order $\epsilon^\prime/\epsilon$.
Furthermore, because of the small phase space for
$K\rightarrow 3\pi$ decays or because one is dealing with electromagnetic
rescattering in $K\rightarrow \pi\pi\gamma$, the rescattering
contribution suppress these asymmetries even more.  Table 2 gives
typical predictions, contrasting them to the expected reach of
the Frascati $\Phi$ factory with $\int {\cal{L}}~dt=10~{\rm fb}^{-1}$.
For the $K\rightarrow 3\pi$ decays, Belkov {\it et al.} \cite{Belkov}
give numbers at least a factor of 10 above those given in Table 2.
However, these numbers are predicated on having very large rescattering
phases which do not appear to be realistic\cite{IMP}.  One is lead to
conclude that, if the CKM paradigm is correct, it is unlikely that
one will see a CP-violating signal in charged Kaon decays.

\begin{table}
\caption{Predictions for Asymmetries in $K^\pm$ Decays}
\begin{center}
\begin{tabular}{|c|c|c|} \hline
Asymmetry & Prediction & $\Phi$ Factory Reach \\ \hline
${\cal{A}}(\pi^+\pi^+\pi^-;\pi^-\pi^-\pi^+)$ &
$5\times 10^{-8}~~~\mbox{\cite{Pettit}}$ & $3\times 10^{-5}$ \\
${\cal{A}}(\pi^+\pi^o\pi^o;\pi^-\pi^o\pi^o)$  &
$2\times 10^{-7}~~~\mbox{\cite{Pettit}}$ & $5\times 10^{-5}$ \\
${\cal{A}}_{\rm Dalitz}(\pi^+\pi^+\pi^-;\pi^-\pi^+\pi^+)$ &
$2\times 10^{-6}~~~\mbox{\cite{Pettit}}$ & $3\times 10^{-4}$ \\
${\cal{A}}_{\rm Dalitz}(\pi^+\pi^o\pi^o;\pi^-\pi^o\pi^o)$ &
$1\times 10^{-6}~~~\mbox{\cite{Pettit}}$ & $2\times 10^{-4}$ \\
${\cal{A}}(\pi^+\pi^o\gamma;\pi^-\pi^o\gamma)$ &
$10^{-5}~~~\mbox{\cite{HYC}}$ & $2\times 10^{-3}$ \\  \hline
\end{tabular}
\end{center}
\end{table}

\subsection*{$K_{\rm L}\rightarrow \pi^o\ell^+\ell^-;~K_{\rm L}
\rightarrow \pi^o\nu \bar\nu$}

Perhaps more promising are decays of the $K_{\rm L}$ to $\pi^o$ plus
lepton pairs.  If the lepton pair is charged, then the process has
a CP conserving piece in which the decay proceeds via a $2\gamma$
intermediate state.  Although there was some initial
controversy \cite{Seghal}, the rate for the process $K_{\rm L}\rightarrow
\pi^o\ell^+\ell^-$ arising from the CP-conserving $2\gamma$ transitiion
is probably at, or below, the $10^{-12}$ level \cite{Dan}:
\[
B(K_{\rm L}\rightarrow \pi^o\ell^+\ell^-)_{\rm CP~cons.}=
(0.3-1.2)\times 10^{-12}
\]
and is just a small correction to the dominant CP violating
contribution going through an effective spin 1 virtual state,
$K_{\rm L}\rightarrow \pi^oJ^*$.  Since $\pi^o J^*$ is CP even,
this part of the amplitude is CP violating and
has two distinct pieces \cite{Dib}: an
indirect contribution from the CP even piece ($\epsilon K_1$) in the
$K_{\rm L}$ state and a direct $\Delta S=1$ CP-violating
piece coming from the $K_2$ part of $K_L$:
\[
A(K_{\rm L}\rightarrow \pi^o J^*)=\epsilon A(K_1\rightarrow \pi^o J^*)+
A(K_2\rightarrow \pi^o J^*)~.
\]

To isolate the interesting direct CP contribution in this process
requires understanding first the size of the indirect contribution.
The amplitude $A(K_1\rightarrow \pi^o J^*)$ could be determined
absolutely if one had a measurement of the process
$K_{\rm S}\rightarrow \pi^o\ell^+\ell^-$.  Since this is not at
hand, at the moment one has to rely on various guesstimates.
These give the following range for the indirect CP-violating
branching ratio\cite{WW}
\[
B(K_{\rm L}\rightarrow \pi^o\ell^+\ell^-)^{\rm indirect}_{\rm CP~violating}=
(1.6-6)\times 10^{-12}~,
\]
where the smaller number is the estimate coming from chiral
perturbation theory, which the other comes from
relating $A(K_1\rightarrow \pi^o J^*)$ to the analogous amplitude
for charged K decays.

The calculation of the direct CP-violating contribution to the
process $K_{\rm L}\rightarrow \pi^o\ell^+\ell^-$, as a result of
electroweak Penguin and box contributions and their gluonic
corrections, is perhaps the one that is most reliably known.  The
branching ratio obtained by Buras, Lautenbacher, Misiak and M\"unz in
their next to leading order calculation
of the process\cite{BLMM} is
\[
B(K_{\rm L}\rightarrow \pi^o\ell^+\ell^-)^{\rm direct}_{\rm CP~violating}=
(5\pm 2)\times 10^{-12}
\]
where the error arises mostly from the imperfect knowledge of the CKM
matrix.

Experimentally one has the following 90\% C.L. for the two
$K_{\rm L}\rightarrow \pi^o\ell^+\ell^-$ processes:
\begin{eqnarray*}
B(K_{\rm L}\rightarrow \pi^o\mu^+\mu^-) & < & 5.1\times 10^{-9} \\
B(K_{\rm L}\rightarrow \pi^o e^+e^-) & < & 1.8\times 10^{-9}
\end{eqnarray*}
The first limit comes from the E799 experiment at Fermilab\cite{Harris},
while the second limit combines the bounds obtained by the E845
experiment at Brookhaven\cite{Ohl} and the E799 Fermilab
experiment\cite{DHH}.  Forthcoming experiments at KEK and Fermilab
should be able to improve these limits by at least an order of
magnitude\footnote{KEK 162 goal is to get to a BR of $O(10^{-10})$
while KTeV hopes to push this BR down to $5\times 10^{-11}$.}, if they
can control the dangerous background arising from the decay
$K_{\rm L}\rightarrow \gamma\gamma e^+e^-$\cite{Greenlee}.  Even more
distant future experiment may actually reach the level expected
theoretically for the $K_{\rm L}\rightarrow \pi^o e^+e^-$
rate \cite{WINS}.
However, it will be difficult to unravel the direct CP-violating
contribution from the indirect CP-violating contribution, unless
the $K_{\rm S}\rightarrow \pi^oe^+e^-$ rate is also measured
simultaneously.

In this respect, the process $K_{\rm L}\rightarrow \pi^o\nu \bar\nu$
is very much cleaner.  This process is purely CP-violating, since it
has no $2\gamma$ contribution.  Furthermore, it has a tiny indirect CP
contribution, since this is of order $\epsilon$ times the already
small $K^+\rightarrow \pi^+\nu\bar\nu$ amplitude\cite{Littenberg}.
Next to leading QCD calculations for the direct rate have been
carried out by Buchalla and Buras\cite{BBB} who give the following
approximate formula for the branching ratio for this process
\[
B(K_{\rm L}\rightarrow \pi^o\nu\bar\nu)=8.2\times 10^{-11}
A^4\eta^2\left(\frac{m_t}{M_{\rm W}}\right)^{2.3}~.
\]
This value is very far below the present 90\% C.L. obtained by
the E799 experiment at Fermilab\cite{Weaver}
\[
B(K_{\rm L}\rightarrow \pi^o\nu\bar\nu)<5.8\times 10^{-5}~.
\]
KTeV should be able to lower this bound substantially, perhaps
to the level of $10^{-8}$ but this still leaves a long way to
go!

\subsection*{$K^+\rightarrow \pi^+\nu\bar\nu$}

The last process I would like to consider is the charged Kaon analogue
to the process just discussed.  Although the decay
$K^+\rightarrow \pi^+\nu\bar\nu$ is not CP violating, it is
sensitive to $|V_{\rm td}|^2 \simeq A^2\lambda^6[(1-\rho)^2+\eta^2]$
and so, indirectly, it is sensitive to the CP violating CKM
parameter $\eta$.
For the CP violating decay $K_{\rm L}\rightarrow \pi^o\nu\bar\nu$
the electroweak Penguin and box contributions are dominated by loops
containing top quarks.  Here, because one is not looking at the
imaginary part one cannot neglect altogether the contribution from charm
quarks.  If one could do so, the branching ratio formula for
$K^+\rightarrow \pi^+\nu\bar\nu$ would be given by an analogous
formula to that for $K_{\rm L}\rightarrow \pi^o\nu\bar\nu$ but with
$\eta^2\rightarrow \eta^2+(1-\rho)^2$.

Because $m_t$ is large, the $K^+\rightarrow \pi^+\nu\bar\nu$
branching ratio is not extremely sensitive to the contribution of
the charm-quark loops \cite{Dibc}.  Furthermore, when next to leading
QCD corrections are computed the sensitivity of the result to the
precise value of the charm-quark mass is reduced considerably\cite{BB2}.
Buras {\it et al.}\cite{waiting} give the following approximate
formula for the $K^+\rightarrow \pi^+\nu\bar\nu$ branching ratio
\[
B(K^+\rightarrow \pi^+\nu\bar\nu)=2\times 10^{-11}
A^4\left[\eta^2+\frac{2}{3}(\rho^e-\rho)^2+\frac{1}{3}
(\rho^\tau-\rho)^2\right]
\left(\frac{m_t}{M_{\rm W}}\right)^{2.3}~.
\]
In the above the parameters $\rho^e$ and $\rho^\tau$ differ from
unity because of the presence of the charm-quark contributions.  Taking
$m_t=175~{\rm GeV}$ and $m_c(m_c)=1.30 \pm 0.05~{\rm GeV}$ \cite{GL},
Buras {\it et al.}\cite{waiting} find that $\rho^e$ and $\rho^\tau$
lie in the ranges
\[
1.42 \leq \rho^e \leq 1.55~; \;\; 1.27 \leq \rho^\tau \leq 1.38~.
\]

Using the range of $\eta$ and $\rho$ determined by the CKM
analysis discussed here
gives about a 40\% uncertainty for the $K^+\rightarrow \pi^+\nu\bar\nu$
branching ratio, leading to the expectation
\[
B(K^+\rightarrow \pi^+\nu\bar\nu)=(1\pm 0.4)\times 10^{-10}~.
\]
This number is to be compared to the best present limit coming
from the E787 experiment at Brookhaven.  Careful cuts must be made
in the accepted $\pi^+$ range and $\pi^+$ momentum to avoid
potentially dangerous backgrounds, like $K^+\rightarrow \pi^+\pi^o$
and $K^+\rightarrow \mu^+\pi^o\nu$.  Littenberg\cite{LINS} at this
meeting has given a new preliminary result for this branching
ratio
\[
B(K^+\rightarrow \pi^+\nu\bar\nu)<3\times 10^{-9}~~~
\mbox{(90\% C.L.)}
\]
which updates the previously published result from the E787
collaboration\cite{Atiyah}.  This value is still about a factor
of 30 from the interesting CKM model range, but there are hopes
that one can get close to this sensitivity in the present run of
this experiment.

\subsubsection*{Looking for new CP-violating phases}

Positive signals for $\epsilon^\prime/\epsilon \not= 0$ will indicate
the general validity of the CKM picture.  However, given the large
theoretical uncertainty, it is clear that values of $\epsilon^\prime/
\epsilon$ consistent with zero at the $10^{-4}$ level cannot disprove
this picture.  In my view, it is more likely that B-physics
experiments (particularly the detection of the expected large
asymmetry in $B_d\rightarrow \psi K_{\rm S}$ decays\cite {Gronau}) will
provide the crucial smoking gun for the CKM paradigm, with rare Kaon
decays filling in the detailed picture.  However, whether the CKM
picture is (essentially) correct or not, experiments in the Kaon
sector may provide the first glimpse at {\bf other} CP-violating
phases.

There are good theoretical arguments for having further CP-violating
phases, besides the CKM phase $\delta$.  For instance, to establish
a matter-antimatter asymmetry in the Universe one needs to have
processes which involve CP violation\cite{Sakharov}.  If the origin
of this asymmetry comes from processes at the GUT scale, then, in
general, the GUT interactions contain further CP-violating phases besides
the CKM phase $\delta~$\cite{PecceiB}.  If this asymmetry is
established at the electroweak scale\cite{Shap}, then most likely
one again needs further phases, both because intrafamily suppression
gives not enough CP violation in the CKM case to generate the asymmetry
and because one needs to
have more than one Higgs doublet\cite{Cohen}.  Indeed this last point
gives the fundamental reason why one should expect to have further
CP-violating phases, besides the CKM phase $\delta$.  It is likely
that the standard model is part of a larger theory.  For instance,
supersymmetric extensions of the SM have been much in vogue
lately.  Any such extensions will introduce further particles and
couplings and thus, by the simple corollary mentioned at the beginning
of this section, they will introduce new CP-violating phases.

The best place to look for non-CKM phases is in processes where
CP violation within the CKM paradigm is either vanishing or very
suppressed.  One such example is provided by experiments aimed at
measuring the electric dipole moments of the neutron or the electron,
since electric dipole moments are predicted to be extremely small
in the CKM model.  Another example concerns measurements of the
transverse muon polarization $\langle P_\perp^\mu\rangle$ in
$K_{\mu 3}$ decays, which vanishes in the CKM paradigm\cite{Leurer}.
The transverse muon polarization measures a T-violating triple
correlation\cite{Sakurai}
\[
\langle P_\perp^\mu\rangle \sim \langle
\vec s_\mu \cdot (\vec p_\mu\times \vec p_\pi)\rangle~.
\]
In as much as one can produce such an effect also as a result of
final state interactions (FSI) this is not a totally clear test
for new CP-violating phases.  With two charged particles in the
final state, like for the decay $K_{\rm L}\rightarrow \pi^-\mu^+\nu_\mu$,
one expects FSI to give typically
$\langle P_\perp^\mu\rangle_{\rm FSI} \sim \alpha/\pi \sim
10^{-3}$~\cite{Adkins}.  However, for the process
$K^+\rightarrow \pi^o\mu^+\nu_\mu$ with only one charged particle
in the final state, the FSI effects should be much smaller.  Indeed,
Zhitnitski\cite{Zhitnitski} estimates for this proceses that
$\langle P_\perp^\mu\rangle_{\rm FSI}\sim 10^{-6}$.  So
a $\langle P_\perp^\mu\rangle$ measurement in the $K^+\rightarrow
\pi^o\mu^+\nu_\mu$ decay is a good place to test for additional
CP-violating phases.

The transverse muon polarization $\langle P_\perp^\mu\rangle$ is
particularly sensitive to scalar interactions and thus is a good
probe of CP-violating phases arising from the Higgs
sector\cite{CFK}.  One can write the effective $K_{\mu 3}$
amplitude\cite{BG} as
\[
A=G_{\rm F}\sin\theta_c f_+(q^2)
\left\{p_K^\mu \bar\mu \gamma_\mu(1-\gamma_5)
\nu_\mu + f_S(q^2)m_\mu \bar\mu(1-\gamma_5)^\nu_\mu\right\}~.
\]
Then
\[
\langle P_\perp^\mu\rangle = \frac{m_\mu}{M_{\rm K}} {\rm Im}~f_{\rm S}
\left[\frac{|\vec p_\mu|}
{E_\mu+|\vec p_\mu|n_\mu \cdot n_\nu-m_\mu^2/M_{\rm K}}\right]
\simeq 0.2~{\rm Im}~f_{\rm S}~.
\]
Here $n_\nu(n_\nu)$ are unit vectors along the muon (neutrino)
directions and the numerical value represents the expectation after
doing an average over phase space\cite{Kuno}.

Contributions to ${\rm Im}~f_{\rm S}$ can arise in multi-Higgs models
(like the Weinberg 3-Higgs model\cite{Weinberg}) from the charged
Higgs exchange shown in Fig. 5, leading to \cite{Higgs}
\[
{\rm Im}~f_{\rm S} \simeq {\rm Im}(\alpha^*\gamma)
\frac{M_{\rm K}^2}{M_{{\rm H}^-}^2}~.
\]
Here $\alpha(\gamma)$ are constants associated with the charged
Higgs coupling to quarks (leptons).  Because a leptonic vertex
is involved, one in general does not have a strong constraint on
${\rm Im}(\alpha^*\gamma)$.  By examining possible non-standard
contributions to the B semileptonic decay $B\rightarrow X\tau \nu_\tau$,
Grossman\cite{Grossman} obtains
\[
{\rm Im}(\alpha^*\gamma) <
\frac{0.23~M^2_{H^-}}{[{\rm GeV}]^2}
\]
which yields a bound for $\langle P_\perp^\mu\rangle$ of
$\langle P_\perp^\mu\rangle < 10^{-2}$.  Amusingly, this is
the same bound one infers from the most accurate measurement of
$\langle P_\perp^\mu\rangle$ done at Brookhaven about a decade
ago \cite{Blatt}, which yielded
\[
\langle P_\perp^\mu\rangle = (-3.1\pm 5.3)\times 10^{-3}~.\]
\begin{figure}[h!]
{}~\epsfig{file=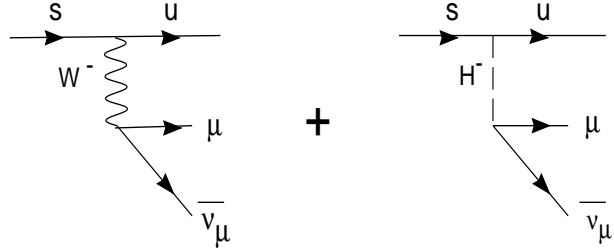,width=13.5cm,height=3.0cm}
\caption{Graphs contributing to $\langle P_\perp^\mu\rangle$}
\end{figure}

In specific models, however, the leptonic phases associated with
charged Higgs couplings are correlated with the hadronic phases.
In this case,
one can obtain more specific restrictions on $\langle P_\perp^\mu\rangle$
due to the strong bounds on the neutron electric dipole moment.
For instance, for the Weinberg 3 Higgs model, one relates
${\rm Im} (\alpha^*\gamma)$ to a similar product of couplings of
the charged Higgs to quarks\cite{Higgs}:
\[
{\rm Im}(\alpha^*\gamma)=\left(\frac{v_u}{v_e}\right)^2~
{\rm Im}(\alpha^*\beta)~,
\]
where $v_u~(v_e)$ are the VEV of the Higgs doublets which couples
to up-like quarks (leptons).  The strong bound on the neutron electric
dipole moment\cite{PDG} then gives the constraint
\[
{\rm Im}(\alpha^*\beta) \leq
\frac{4\times 10^{-3}~M_{{\rm H}^-}^2}{[{\rm GeV}]^2}~.
\]
If one assumes that $v_u \sim v_e$, this latter bound gives
a strong constraint on $\langle P_\perp^\mu\rangle \break
[\langle P_\perp^\mu\rangle < 10^{-4}]$.
However, this constraint is removed if
$v_u/v_e \sim m_t/m_\tau$.

Similar results are obtained in the simplest supersymmetric
extension of the SM.  In this case, ${\rm Im}~f_{\rm S}$ arises from
a complex phase associated with the gluino mass.  Assuming all
supersymmetric masses are of the same order, Christova and
Fabbrichesi\cite{CF} arrive at the estimate
\[
{\rm Im}~f_{\rm S} \simeq \frac{M_{\rm K}^2}{m_{\tilde g}^2}
\frac{\alpha_{\rm s}}{12\pi} \sin\phi_{\rm susy}~,
\]
where $\phi_{\rm susy}$ is the gluino mass CP-violating phase.
This phase, however, is strongly restricted by the neutron electric
dipole moment.  Typically, one finds\cite{Hall}
\[
\sin\phi_{\rm susy} \leq \frac{10^{-7}~m^2_{\tilde g}}
{[{\rm GeV}]^2}
\]
leading to a negligible contribution for
$\langle P_\perp^\mu\rangle$,  below the level of
$\langle P_\perp^\mu\rangle_{\rm FSI}$.

An experiment (E246) is presently underway at KEK aimed at
improving the bound on $\langle P_\perp^\mu\rangle$ obtained
earlier at Brookhaven.  The sensitivity of E246 is such that one should
be able to achieve an error $\delta\langle P_\perp^\mu\rangle \sim
5\times 10^{-4}$\cite{Kuno}.  This level of precisiion is very
interesting and, in some ways, it is comparable or better to
$d_n$ measurements for probing CP-violating phases from the scalar
sector.  This is the case, for instance, in the Weinberg model if
$v_u/v_e$ is large.  At any rate, if a positive signal were
to be found, it would be a clear indication for a non-CKM CP-violating
phase.  Furthermore, as Garisto\cite{Garisto} has pointed out, a
positive signal at the level aimed by the E246 experiment would
imply very large effects in the corresponding decays in the B system
involving $\tau$-leptons (processes like $B^+\rightarrow D^o \tau^+
\nu_\tau$), since one expects, roughly,
\[
\langle P_\perp^\tau\rangle_{\rm B} \sim
\frac{M_{\rm B}}{M_{\rm K}} \frac{m_\tau}{m_\mu}
\langle P_\perp^\mu\rangle_{\rm K}~.
\]
Thus, in principle, a very interesting experimental cross-check
could be done.

\section{Concluding Remarks}

In the past we have learned profound lessons by doing experiments
with Kaon beams.  It is my impression that in the future
we will continue to
learn from Kaons important information,
as the planned experiments
have an increasing level of precision and sophistication.
Indeed, in the next five years, there are a number of experiments
which could produce big surprises [flavor violation; CPT
violation; evidence for non-CKM phases; decay rates above the SM
expectations] and others which could further strengthen our
present paradigm for CP violation, through a non-zero measurement
of $\epsilon^\prime/\epsilon$.

This said, it is a fact that all the experiments presently under
construction or taking data are extraordinarily hard and require
tremendous sophistication.  Thus it seems almost inconceivable
(impossible?) to go beyond them.  For this reason, it would seem
sensible to me to adopt a ``plan now, decide later" attitude for new
Kaon experiments, beyond those now on the books.  That is, it would
seem prudent before deciding to go the next step to await the results
of the data which will be forthcoming in the next half decade.

\section*{Acknowledgments}

I would like to thank Professor S. T. Yamazaki and S. Yamada for their
hospitality at the INS Symposium.
This work was supported in part by the
Department of Energy under Grant No. FG03-91ER40662.

\vspace{1pc}

\end{document}